\documentclass[10pt, conference]{IEEEtran}
\IEEEoverridecommandlockouts

\usepackage{graphicx}
\usepackage{subfigure}
\usepackage{amsmath}
\usepackage{amsfonts}
\usepackage{mathrsfs}
\usepackage[linesnumbered,ruled,vlined]{algorithm2e} 
\usepackage{algpseudocode}

\usepackage{threeparttable}
\usepackage{url}

\usepackage[absolute]{textpos}
\begin{document}

\begin{textblock}{18}(2,0.5)
    \centering
    \noindent X. Zhang, S. Sun, M. Tao, Q. Huang, and X. Tang, “Joint hybrid beamforming and user scheduling for multi-satellite
    cooperative networks,” accepted to \textit{the 2023 IEEE Wireless Communications and Networking Conference (WCNC)}, Glasgow, Scotland, UK, Mar. 2023.
 \end{textblock}

\title{Joint Hybrid Beamforming and User Scheduling\\ for Multi-Satellite Cooperative Networks}


\author{\IEEEauthorblockN{Xuan Zhang\IEEEauthorrefmark{1},
Shu Sun\IEEEauthorrefmark{1},
Meixia Tao\IEEEauthorrefmark{1}, 
Qin Huang\IEEEauthorrefmark{2} and
Xiaohu Tang\IEEEauthorrefmark{3}}
\IEEEauthorblockA{\IEEEauthorrefmark{1}Department of Electronic Engineering, Shanghai Jiao Tong University, Shanghai, China\\}
\IEEEauthorblockA{\IEEEauthorrefmark{2}School of Electronic and Information Engineering, Beihang University, Beijing, China\\}
\IEEEauthorblockA{\IEEEauthorrefmark{3}School of Information Science and Technology, Southwest Jiaotong University, Chengdu, Sichuan, China\\}
Emails: \IEEEauthorrefmark{1}\{zhangxuan1998, shusun, mxtao\}@sjtu.edu.cn, \IEEEauthorrefmark{2}qhuang.smash@gmail.com, \IEEEauthorrefmark{3}xhutang@swjtu.edu.cn

\thanks{This work is supported by the National Natural Science Foundation of China under grant 61941106.}
}


%


\maketitle

\begin{abstract}
In this paper, we consider a cooperative communication network where multiple satellites provide services for ground users (GUs) (at the same time and on the same frequency). 
The communication and computational resources on satellites are usually restricted and the satellite-GU link determination affects the communication performance significantly when multiple satellites provide services for multiple GUs in a collaborative manner.
Therefore, considering the limitation of the on-board radio-frequency chains, we first propose a hybrid beamforming method consisting of analog beamforming for beam alignment and digital beamforming for interference mitigation. 
Then, to establish appropriate connections between satellites and GUs, we propose a heuristic user scheduling algorithm which determines the connections according to the total spectral efficiency (SE) increment of the multi-satellite cooperative network. 
Next, a joint hybrid beamforming and user scheduling scheme is proposed to dramatically improve the performance of the multi-satellite cooperative network. 
Moreover, simulations are conducted to compare the proposed schemes with representative baselines and analyze the key factors influencing the performance of the multi-satellite cooperative network. It is shown that the proposed joint beamforming and user scheduling approach can provide 47.2\% SE improvement on average as compared with its non-joint counterpart.
\end{abstract}

\begin{IEEEkeywords}
Satellite communication, hybrid beamforming, user scheduling.
\end{IEEEkeywords}

%
\IEEEpeerreviewmaketitle

\section{Introduction}
Satellites have a distinctive ability of covering wide geographical areas through a minimum amount of infrastructure on the ground.
Currently, the field of satellite communications is drawing increasing attention in the global telecommunications market. Several network operators start using satellites in backhaul infrastructures for connectivity and for 5G system integration \cite{5G}. 
Furthermore, satellites are likely to play an increasingly important role in the 6G era to provide global seamless coverage and support space-terrestrial integrated networks. 
The integrated architectures, applications, and challenges of satellite-terrestrial networks toward 6G are presented in \cite{integrated-6G}.

In \cite{Security}, the authors introduced several satellite communication networks which are categorized by different architectures, including land mobile satellite communication networks, hybrid satellite-terrestrial relay networks, and satellite-terrestrial integrated networks. 
The authors in \cite{leo-comparison} made a technical comparison of three low-Earth-orbit (LEO) satellite constellation systems: OneWeb's, SpaceX's, and Telesat's. The advantages and technical challenges of  multi-satellite cooperative transmission systems in 5G were discussed in \cite{multi-sat}. 

In satellite communication systems, due to the limitation of the on-board radio-frequency (RF) chains caused by the confined device complexity and transmission power, hybrid analog and digital beamforming is a promising method to balance performances and hardware constraints. In \cite{hybrid}, a hybrid beamforming strategy was used for massive multiple-input multiple-output LEO satellite communications. Later, the authors in \cite{hybrid-RIS} investigated hybrid beamforming method for reconfigurable intelligent surface-assisted secure integrated terrestrial-aerial networks. 
Besides, user scheduling is another factor that affects the performances of satellite communication networks. 
In \cite{us1}, an adaptive user scheduling method was proposed to mitigate intra-beam and inter-beam interference under the single-satellite scenario.
Nevertheless, cooperation among multiple satellites was not taken into account in \cite{hybrid}--\cite{us1}.
The authors in \cite{us2} proposed a multilevel clustering algorithm and a cross-cluster grouping algorithm to realize user scheduling, which considered the multi-satellite cooperation, but the user scheduling was only dependent on the positions of users. 
Furthermore, full frequency reuse (FFR) is widely adopted 
to improve the spectral efficiency (SE). However, aggressive frequency reuse will inevitably cause severe inter-beam interference.
As such, multi-satellite collaboration via cooperative digital beamforming is crucial for inter-beam interference mitigation. The authors in \cite{dig-beamf} compared the performances of several common digital beamforming designs: zero-forcing (ZF), regularized ZF and MMSE digital beamformer. 
In \cite{us1} and \cite{joint}, beamforming design and resource allocation were considered for the integrated terrestrial-satellite network. The authors in \cite{coexist} studied the transmit beamforming design for spectral coexistence of satellite and terrestrial networks. 
However, to our best knowledge, there is little existing work on the joint design of beamforming and user scheduling for multi-satellite cooperative networks without the assistance of terrestrial systems.

In this paper, we investigate beamforming and user scheduling in a multi-satellite cooperative network with FFR. Considering the limitation of on-board device complexity and transmission power, we propose a hybrid beamforming method based on the hybrid architecture of satellite antennas. It consists of analog beamforming based on a discrete Fourier transform (DFT) codebook for beam alignment and digital beamforming for interference mitigation. The user scheduling method is also important when multiple satellites work cooperatively. To reasonably arrange multiple satellites to provide services for ground users (GUs), we propose a low-complexity heuristic user scheduling algorithm. Considering the intrinsic connection between beamforming and user scheduling, we propose a joint hybrid beamforming and user scheduling scheme for maximizing the network SE. Simulation results demonstrate that the proposed joint scheme outperforms a representative non-joint baseline strategy dramatically and can achieve 47.2\% SE increase on average.

\section{System Model}
\subsection{System Architecture} \label{sec-2A}
\begin{figure}[tb]
    \centering
    \includegraphics[width=0.48\textwidth]{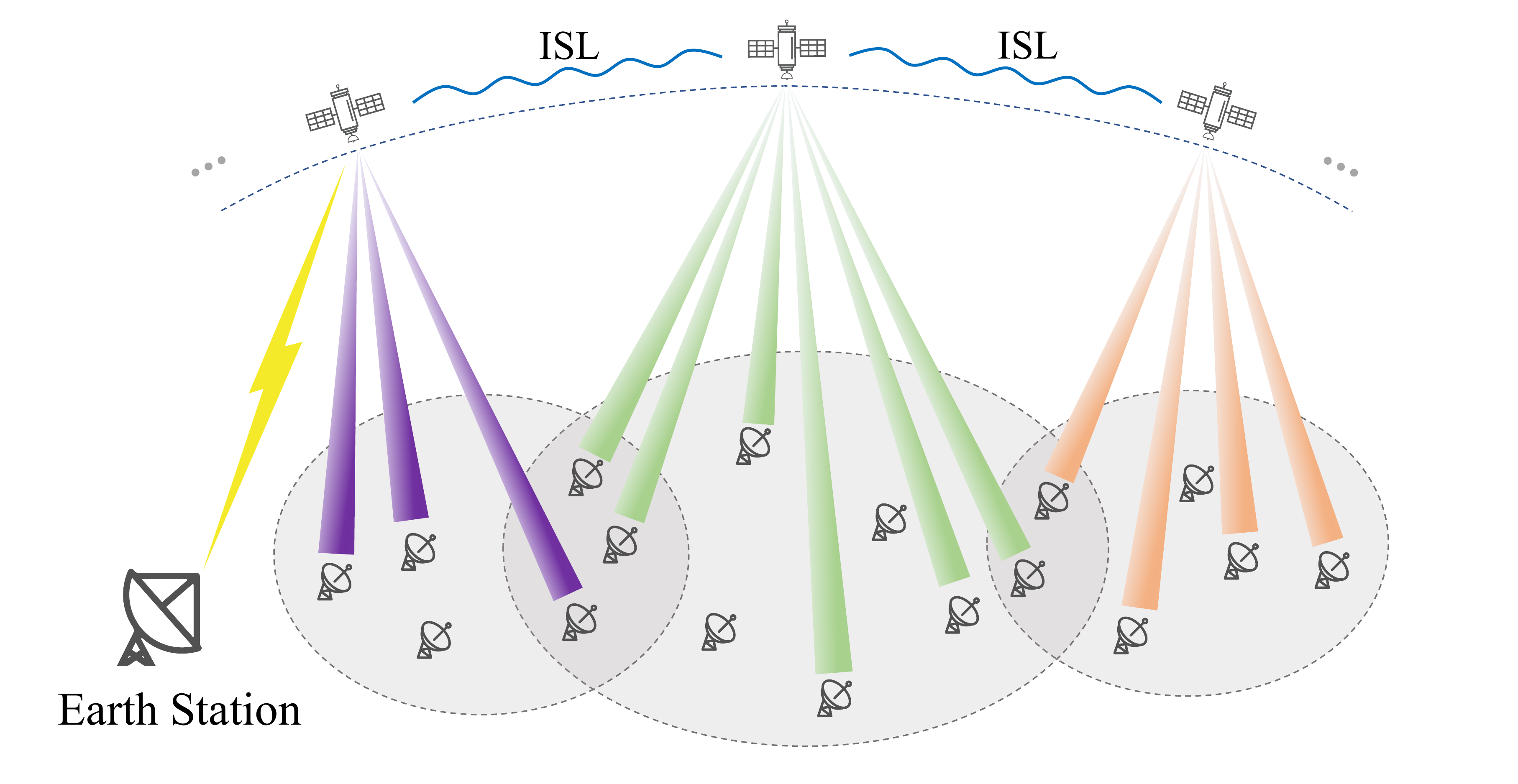}
    \caption{System architecture of the multi-satellite cooperative communication network in this work. ISL denotes inter-satellite link.}
    \label{system}
    \vspace{-0.3cm}
\end{figure}

As depicted in Fig. \ref{system}, we consider a downlink communication scenario of a multi-satellite communication network at a certain moment, where multiple satellites cooperatively provide services for GUs. We assume that there are $N_u$ GUs requesting services and $N_s$ satellites visible to at least one of these GUs. 
We use $\mathcal{V}_g$ to denote the set of visible satellites of GU $g$ and $\mathcal{V}_s$ for the set of GUs that satellite $s$ is serving. We also assume that all the satellites are equipped with regenerative payload and belong to an LEO constellation operating in the Ka-band with FFR adopted. Furthermore, there are optical inter-satellite links (ISLs) to exchange data among satellites, and the satellites can perform on-board distributed computing to share computation load \cite{distributed-computing-1}. 
The earth station periodically sends topological relationships of the constellation to the satellites via its line of sight (LoS) and satellites can share the topological relationships through ISLs.

Consider that each GU is equipped with a very small aperture terminal (VSAT) which is a single antenna system, and each satellite is equipped with a uniform planner array (UPA). Different from purely analog or digital beamforming architecture, the UPA here adopts a hybrid architecture. 
The UPA is composed of $N_b = N_{x}^\text{sub} \times N_{y}^\text{sub}$ sub-arrays where $x$ denotes the axis pointing in the direction of the satellite's movement and $y$ denotes the axis pointing in the direction orthogonal to the satellite's movement. Each sub-array consists of $N = N_x \times N_y$ antenna elements, and each sub-array is connected with one RF chain, generating one independent spot beam. Therefore, one satellite can generate $N_b$ independent spot beams at most. Besides, we assume that each independent beam serves one GU. This indicates that one satellite can provide services for up to $N_b$ GUs simultaneously.

\subsection{Channel Model}
We consider the scenario where the satellite-GU link is under the condition of LoS and clear sky (no rain and cloud attenuation), and all the GUs are distributed in suburban areas. The channel is modeled according to the technical reports of 3GPP \cite{3GPP-38811} and ITU-R \cite{ITU-R-681}. The multiple-input single-output (MISO) channel between satellite $s$, for $s \in \{1,2,\dots,N_s\}$ and GU $g$ for $g \in \{1,2,\dots,N_u\}$ can be modeled as
\begin{equation}  \label{eq-channel}
    \textbf{h}_{sg} = \xi_{sg} \cdot \textbf{h}_{sg_{s}},
\end{equation}
where $\xi_{sg}$ and $\textbf{h}_{sg_{s}} \in \mathbb{C}^{N \times 1}$ stand for the radio propagation loss and the small-scale fading channel between satellite $s$ and GU $g$, respectively. Here, $\xi_{sg}$ is directly relevant to the large-scale path loss (PL). The PL can be calculated as follows:
\begin{equation}
    \text{PL}[\text{dB}] = \text{PL}_\text{b}[\text{dB}] + \text{PL}_\text{g}[\text{dB}] + \text{PL}_\text{s}[\text{dB}],
\end{equation}
where $\text{PL}_\text{b}$ denotes the basic path loss, including the distance- and frequency-dependent free space path loss and the log-normal distributed shadow fading, $\text{PL}_\text{g}$ denotes the attenuation due to atmospheric gasses, and $\text{PL}_\text{s}$ denotes the attenuation due to tropospheric scintillation.

The small-scale fading channel model follows a Loo distribution, where the received signal is the sum of two components: the direct path and the diffuse multipath. We assume that there are $N_\text{cl}$ clusters with $N_\text{ray}$ propagation paths in each cluster. The small-scale fading channel $\textbf{h}_{sg_s}$ can be modeled as
\begin{equation}
    \begin{split}
        \textbf{h}_{sg_s} = \delta & \bigg( m_0 \, \textbf{a}_\text{T}(\phi_{sg}, \theta_{sg}) 
         + \sum_{l=1}^{N_\text{cl}} \sum_{i=1}^{N_\text{ray}} m_{li} \textbf{a}_\text{T}(\phi_{sg,li}, \theta_{sg,li}) \bigg),
    \end{split}
\end{equation}
where $m_0$ and $m_{li}$ are complex coefficients of the direct path and the diffuse multipath, respectively. The amplitude of $m_0$ is subject to the normal distribution, while the amplitude of $m_{li}$ obeys the Rayleigh distribution. The phases of both $m_0$ and $m_{li}$ follow a uniform distribution from 0 to 2$\pi$. The normalization factor $\delta$ is introduced to satisfy $\mathbb{E} \left[ \| \textbf{h}_{sg_s} \|^{2} \right] = 1$. The vector $\textbf{a}_\text{T}(\phi, \theta) \in \mathbb{C}^{N \times 1}$ is the normalized antenna steering vector of the satellite's sub-array, which can be written as
\begin{equation}
    \begin{split}
        \textbf{a}_\text{T}(\phi, \theta) & = \frac{1}{\sqrt{N}} \left[ 1, \dots, e^{-j \frac{2\pi}{\lambda} d \left( p \cos{\theta} \cos{\phi} + q \cos{\theta} \sin{\phi} \right) }, \right. \\
        & \left. \dots,  e^{-j \frac{2\pi}{\lambda} d \left( \left( N_x - 1 \right) \cos{\theta} \cos{\phi} + \left( N_y - 1 \right) \cos{\theta} \sin{\phi} \right) } \right]^T,
    \end{split}
\end{equation}
where $\lambda$ and $d$ are the carrier wavelength and the antenna element spacing, respectively. In our study, we assume $d = \frac{\lambda}{2}$ to guarantee that there is no grating lobe when a beam is steered towards $\pm\, 90^{\circ}$. Besides, $\phi$ and $\theta$ denote the azimuth and elevation angles from the perspective of satellite.

\subsection{Signal Model and Problem Formulation} \label{sec-2C}
As mentioned in Section \ref{sec-2A}, multiple satellites provide communication services for their GUs cooperatively with FFR. Therefore, each GU will suffer interference from all the satellites that are in this GU's LoS. When receiving signals, the GU antenna will aim at the serving satellite and the intended signal and interference from this satellite will experience the maximum gain of the GU antenna. While the interference from other visible satellites will experience a gain decrease due to the off-boresight angles of these satellite-GU links and the narrow-beam characteristics of the VSAT considered herein.
Note that the set of visible satellites of each GU is known, but the specific links between satellites and GUs are unknown and need to be determined. Based on this fact, we introduce a discrete variable $\alpha_{sg}$ to indicate whether the link is established, where $\alpha_{sg} = 1$ if satellite $s$ is providing service for GU $g$, and $\alpha_{sg} = 0$ otherwise.

The received signal of GU $g$ can be expressed as \cite{us1, joint}
\begin{equation} \label{eq-signalmodel}
    y_{g} = \sum_{s \in \mathcal{V}_g} \alpha_{sg} \textbf{h}_{sg}^\text{H} \textbf{w}_{sg} x_{g} + \sum_{g^{'} \neq \, g} \sum_{s \in \mathcal{V}_g} \alpha_{sg^{'}} \textbf{h}_{sg}^\text{H} \textbf{w}_{sg^{'}} x_{g^{'}} + n_g,
\end{equation}
where $\textbf{h}_{sg} \in \mathbb{C}^{N \times 1}$ is the channel vector between satellite $s$ and GU $g$, $\textbf{w}_{sg} \in \mathbb{C}^{N \times 1}$ represents the beamforming vector, and $x_g$ is the requested data of GU $g$, which is assumed to be independent and satisfy $\mathbb{E} \left( |x_g|^2 \right) = 1$. The first term in (\ref{eq-signalmodel}) is the intended data for GU $g$, the second term is the interference coming from the communication services for the other GUs, and the third term is the complex additive white Gaussian noise following the distribution $\mathcal{CN} \left(0, \sigma_s^2 \right)$, where $\sigma_s^2$ denotes the noise power.

Then the received signal-to-interference-plus-noise ratio (SINR) of GU $g$ can be obtained as
\begin{equation}
    \gamma_{g} = \frac{|\sum_{s \in \mathcal{V}_g} \alpha_{sg} \textbf{h}_{sg}^\text{H} \textbf{w}_{sg}|^{2}}
    {\sum_{g^{'} \not = g} | \sum_{s \in \mathcal{V}_g} \alpha_{sg^{'}} \textbf{h}_{sg}^\text{H} \textbf{w}_{sg^{'}}|^{2} + \sigma_{s}^{2}}.
\end{equation}

According to the Shannon theorem, the SE per channel use of GU $g$ can be calculated by
\begin{equation}
    R_{g} = \log_{2}{(1 + \gamma_{g})}.
\end{equation}
As a result, the total SE of the multi-satellite cooperative network can be expressed as
\begin{equation} \label{total-SE}
    R = \sum_{g=1}^{N_u} R_{g}.
\end{equation}

Our objective is to maximize the total SE of the multi-satellite cooperative network by means of calculating the beamforming vector $\textbf{w}_{sg}$ and adjusting the satellite-GU links. Therefore, we formulate the optimization problem $\left(\mathcal{OP} \right)$ as follows:
\begin{align}
    \mathcal{OP}: \quad &
    \max_{\{\textbf{w}_{sg},\alpha_{sg} \}} \qquad \sum_{g=1}^{N_u} \log_{2}{(1 + \gamma_{g})}\\
    s.t.\quad & C_1:\sum_{g} \alpha_{sg} \| \textbf{w}_{sg} \|^{2} \leq P_\text{T}, \, \forall s \in \{1,2,\dots,N_s\},\\
    & C_2:\sum_{g} \alpha_{sg} \leq N_{b}, \, \forall s \in \{1,2,\dots,N_s\},\\
    & C_3:\sum_{s \in \mathcal{V}_g} \alpha_{sg} = 1, \, \forall g \in \{1,2,\dots,N_u\},\\
    & C_4:\alpha_{sg} \in \{ 0,1 \}, \, \forall s,g.
\end{align}
Here, constraint $C_1$ is the power constraint of each satellite, constraint $C_2$ means that the number of GUs connected to the same satellite cannot exceed the maximum number of spot beams that one satellite can generate, namely $N_b$, and constraint $C_3$ is the GU connection constraint. Notably, the $\mathcal{OP}$ is not always feasible because of the coexistence of $C_2$ and $C_3$. When $C_2$ and $C_3$ contradict each other, we give priority to guaranteeing $C_2$ and try to connect as many GUs as possible.
The $\mathcal{OP}$ cannot be solved directly due to the fact that the objective function and constraints are non-convex. Thus, we propose to solve it by the following steps:
\begin{enumerate}
    \item Given $\textbf{h}_{sg}$ in (\ref{eq-channel}), the analog beamforming vector $\textbf{w}_{sg}^\text{A}$ is obtained based on a DFT codebook (with more details to be shown in Algorithm \ref{Codebook-Ana}).
    \item The vector $\textbf{w}_{sg}^\text{A}$ is set as the initial beamforming vector.
    \item Based on 2), we propose two schemes to solve $\mathcal{OP}$, which will be further discussed in the following sections.
\end{enumerate}

\section{Hybrid Beamforming} \label{sec-3}
\subsection{Analog Beamforming Based on Codebook} \label{sec-3A}
We consider analog beamforming for beam alignment based on a codebook, which is widely used in terrestrial systems. Codebook-based beamforming design can reduce the overhead of channel state information (CSI) feedback. In this work, we assume that the GU side has perfect CSI and the GU-satellite CSI feedback link is lossless.
As described in Section \ref{sec-2A}, every satellite is equipped with a UPA, thus the two-dimensional (2D) DFT codebook is applicable and it can be seen as the synthesis of two 1D DFT codebooks in the directions of x and y axes, $\mathbf{D}_x,\mathbf{D}_y$. The $N \times N$ 2D DFT codebook matrix, denoted as $\mathbf{D}$, can be expressed  as

\begin{equation}
    \mathbf{D} = \mathbf{D}_x \otimes \mathbf{D}_y.
\end{equation}

The core idea of analog beamforming herein is to select the best $K$ ($\leq N$) codewords from $\mathbf{D}$ and combine them into a new codeword that satisfies the equal-amplitude constraint of analog beamforming. 
Considering the communication overhead of CSI feedback, $K$ should not be too large and we assume $K = 4$ based on the results of trial simulations. The steps of the proposed analog beamforming method are given in Algorithm \ref{Codebook-Ana}.

\begin{algorithm}[tb] 
    \caption{Codebook-Based Analog Beamforming}
    \label{Codebook-Ana}
    \KwIn{GU-side channel vector $\textbf{h}_{sg},\forall s,g$, codebook $\mathbf{D}$.}
    \KwOut{Analog beamforming vector $\textbf{w}_{sg}^\text{A}$.}
    For each codeword $\mathbf{D}_{:,k}$, calculate $|\textbf{h}_{sg}^\text{H} \mathbf{D}_{:,k}|^2$ and find the best $K$ codewords maximizing it, $\textbf{c}_1,\dots,\textbf{c}_K$\;
    $\mathbf{D}_K = [\textbf{c}_1,\dots,\textbf{c}_K]$, solve the equations $\mathbf{D}_K \textbf{x} = \textbf{h}_{sg}$ and obtain the least square solution $\hat{\textbf{x}} = (\mathbf{D}_K)^{\dagger} \textbf{h}_{sg}$\;
    The GU sends the codeword combination coefficients $\hat{\textbf{x}}$ and codewords' indices to the satellite\;
    Satellite-side combination: $\textbf{w}_{sg}^{'} = \mathbf{D}_K \hat{\textbf{x}}$\;
    \For{$i \in [1, N]$}{$\textbf{w}_{sg}^\text{A}(i) = \textbf{w}_{sg}^{'}(i) \frac{\frac{1}{\sqrt{N}}}{|\textbf{w}_{sg}^{'}(i)|}$}
\end{algorithm}

\subsection{Digital Beamforming}
Based on the link information between satellites and GUs, the channel matrix of satellite $s$ can be written as
\begin{equation} \label{eq-channelmat}
    \mathbf{H}_s = \left[\dots, \textbf{h}_{sg}, \dots \right]^\text{H} \in \mathbb{C}^{N_u^s \times N}, \, g \in \mathcal{V}_s,
\end{equation}
where $N_u^s$ is the number of GUs that satellite $s$ is serving. Similarly, the analog beamforming matrix of satellite $s$ can be written as
\begin{equation} \label{eq-anabeamf}
    \mathbf{F}_s^\text{A} = \left[\dots, \textbf{w}_{sg}^\text{A}, \dots \right] \in \mathbb{C}^{N \times N_u^s}, \, g \in \mathcal{V}_s.
\end{equation}

Hence, we can write the generalized channel matrix between satellite $s$ and the GUs as
\begin{equation}
    \widetilde{\mathbf{H}}_s = \mathbf{H}_s \mathbf{F}_s^\text{A},
\end{equation}
thus the hybrid beamforming is reduced to a digital beamforming problem to mitigate the inter-beam interference of satellite $s$. In this work, we adopt the regularized ZF and the corresponding digital beamforming matrix is
\begin{equation} \label{eq-digbeamf}
    \mathbf{F}_s^\text{D} = \sqrt{\eta} \, \widetilde{\mathbf{H}}_s^\text{H} (\widetilde{\mathbf{H}}_s \widetilde{\mathbf{H}}_s^\text{H} + \beta \mathbf{I}_{N_u^s})^{-1} \in \mathbb{C}^{N_u^s \times N_u^s},
\end{equation}
where $\sqrt{\eta}$ is the power scaling factor to guarantee the satellite is operating at its maximum power, and $\beta$ is an adjustable parameter where $\beta_\text{opt} = \frac{N_u^s \sigma_s^2}{P_\text{T}}$ in the large system limit \cite{ChannelInversion}.

Thus, combining (\ref{eq-anabeamf}) and (\ref{eq-digbeamf}), the hybrid beamforming matrix can be expressed as
\begin{equation} \label{eq-hybeamf}
    \mathbf{F}_s^\text{HY} = \mathbf{F}_s^\text{A} \cdot \mathbf{F}_s^\text{D} = \left[\dots, \textbf{w}_{sg}, \dots \right], \in \mathbb{C}^{N \times N_u^s}, \, g \in \mathcal{V}_s.
\end{equation}

\begin{algorithm}[ht]
    \caption{Heuristic User Scheduling Algorithm }
    \label{PUS}
    \KwIn{Channel matrix $\mathbf{H}_s$, beamforming matrix $\mathbf{F}_s$, and the set of visible satellites $\mathcal{V}_g$, $\forall s \in \{1,2,\dots,N_s\}$, $\forall g \in \{1,2,\dots,N_u\}$.}
    \KwOut{Link matrix $\mathbf{L}$.}
    Initialize $\mathcal{S}=\{1,\dots,N_s\}, \mathcal{G}_n \big|_{n = N_u}=\{1,\dots,N_u\}, \mathbf{L}=\mathbf{0}$\;
    \For{$g \in [1, N_u]$}{
        \If{{\rm length}($\mathcal{V}_g$) $== 1$}{
            $\mathbf{L}(\mathcal{V}_g,g) = 1$\;
            Remove $g$ from $\mathcal{G}_n$ and $n=n-1$\;
        }
    }
    \Repeat{$n==0$}{
        \For{{\rm each possible link} $\mathbf{L}_{sg},\, s \in \mathcal{S}, g \in \mathcal{G}_n$}{
        $\triangle R_{sg} = \mathcal{R}(\mathbf{L}+\mathbf{L}_{sg}, \mathbf{H}_s, \mathbf{F}_s) - \mathcal{R}(\mathbf{L}, \mathbf{H}_s, \mathbf{F}_s)$\;
        }
        $[\hat{s},\hat{g}] = \mathop{\arg\max}\limits_{s,g} \triangle R_{sg}$\;
        \eIf{{\rm satellite} $\hat{s}$ {\rm has spare resource}}{
            $\mathbf{L}(\hat{s},\hat{g}) = 1$\;
            remove $\hat{g}$ from $\mathcal{G}_n$ and $n=n-1$\;
        }{
            remove $\hat{s}$ from $\mathcal{S}$\;
        }
    }
\end{algorithm}

\section{User Scheduling and Implementation Schemes}
\subsection{User Scheduling}
As described in Section \ref{sec-2C}, the links between multiple satellites and GUs need to be determined. The optimal exhaustive search is infeasible here since the computational complexity grows exponentially with the number of GUs. Thus, we propose a heuristic user scheduling algorithm which can achieve a good performance with polynomial complexity as shown in Algorithm \ref{PUS}. 
Therein, $\mathbf{L}$ denotes an $N_s \times N_u$ link matrix where $\alpha_{sg}$ lies in its $s$-th row and $g$-th column, and $\mathbf{L}_{sg}$ denotes the matrix whose $s$-th row and $g$-th column is 1 and other places are 0. The set of satellites that have spare resource is denoted as $\mathcal{S}$. Here, the resource constraint is that the number of GUs that one satellite is serving simultaneously can not exceed $N_b$. The set of unserved GUs is denoted by $\mathcal{G}_n$, where $n$ is the number of GUs in this set. $\triangle R$ indicates the increment of total SE. The total SE is mainly associated with the link matrix $\mathbf{L}$, channel matrix $\mathbf{H}_s$ and the beamforming matrix $\mathbf{F}_s$, which is equivalent to (\ref{total-SE}) and can be abstracted as
\begin{equation}
    R = \mathcal{R}(\mathbf{L}, \mathbf{H}_s, \mathbf{F}_s),
\end{equation}
where $\mathcal{R}$ indicates a function for calculating the total SE.

\begin{figure*}[tb]
    \centering
    \includegraphics[width=0.82\textwidth]{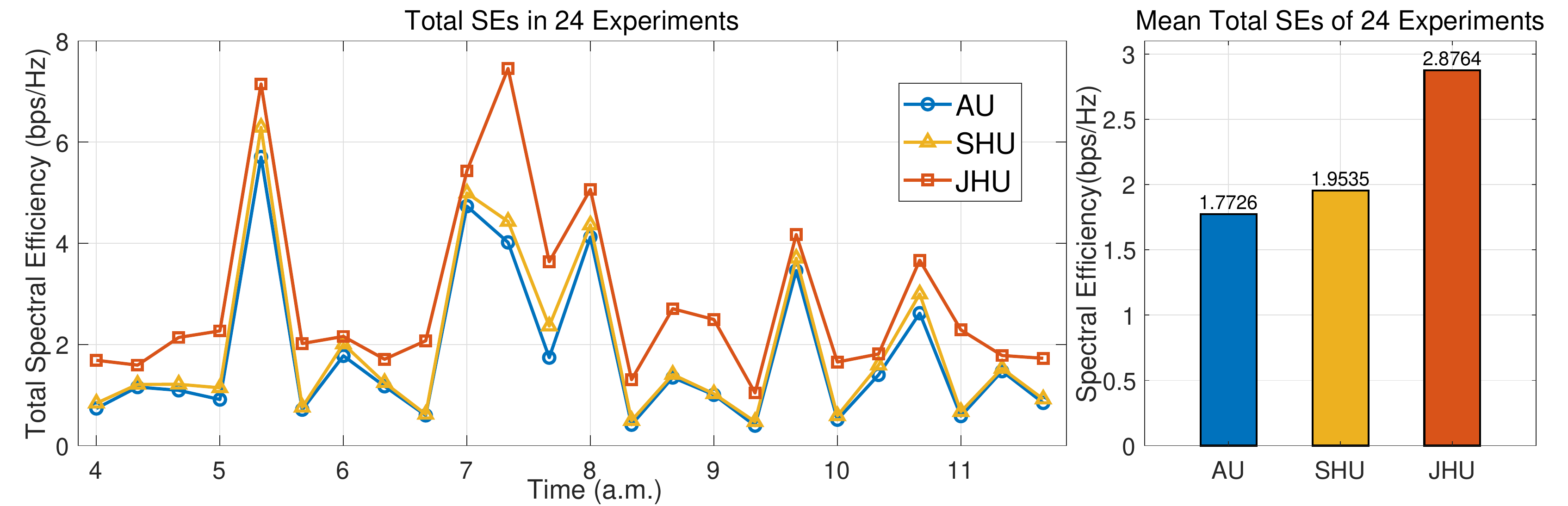}
    \caption{Total SEs and mean total SEs of different schemes.}
    \label{mean-sum}
    \vspace{-0.3cm}
\end{figure*}

\subsection{Separate \& Joint Hybrid Beamforming and User Scheduling Schemes}
In Section \ref{sec-3}, we have introduced the hybrid beamforming method. Within the hybrid beamforming, the analog beamforming can be completed independently. As for the digital beamforming and user scheduling, there are two ways:
\subsubsection{\textbf{\textit{Separate (SHU)}}}
In the SHU scheme, we perform digital beamforming and user scheduling separately and independently. Analog beamforming matrix $\mathbf{F}_s^\text{A}$ is taken as the input of Algorithm \ref{PUS} and user scheduling is performed based on $\mathbf{F}_s^\text{A}$, i.e., $\mathbf{F}_s = \mathbf{F}_s^\text{A}$. Digital beamforming is conducted after user scheduling according to (\ref{eq-digbeamf}), utilizing the final links to mitigate the interference and improve the performance. Finally, we use $\mathbf{F}_s^\text{HY}$ in (\ref{eq-hybeamf}) to calculate the total SE when SHU is adopted.
\subsubsection{\textbf{\textit{Joint (JHU)}}}
The JHU scheme is based on alternating optimization. Different from SHU, user scheduling and digital beamforming in JHU are designed jointly. The beamforming matrix is updated in real time within the user scheduling. As in SHU, $\mathbf{F}_s^\text{A}$ is taken as the initial input of Algorithm \ref{PUS}. The difference is that each time before calculating the SE increment $\triangle R_{sg}$, the hybrid beamforming matrix $\mathbf{F}_s^\text{HY}$ is computed based on the current link matrix. If we abstract (\ref{eq-digbeamf}) and (\ref{eq-hybeamf}) as
\begin{equation}
    \mathbf{F}_s^\text{HY} = \mathcal{F}(\mathbf{L}, \mathbf{H}_s, \mathbf{F}_s^\text{A}),
\end{equation}
where $\mathcal{F}$ is a function for calculating the hybrid beamforming matrix, then Step 8 of Algorithm \ref{PUS} when JHU is adopted can be replaced by
\begin{equation}
\begin{aligned}
    \triangle R_{sg} =\, & \mathcal{R}(\mathbf{L}+\mathbf{L}_{sg}, \mathbf{H}_s, \mathcal{F}(\mathbf{L}+\mathbf{L}_{sg}, \mathbf{H}_s, \mathbf{F}_s^\text{A}))\\
    & - \mathcal{R}(\mathbf{L}, \mathbf{H}_s, \mathcal{F}(\mathbf{L}, \mathbf{H}_s, \mathbf{F}_s^\text{A})).
\end{aligned}
\end{equation}

\section{Performance Evaluation}
\begin{table}[b]
    \vspace{-0.3cm}
    \centering
    \caption{Simulation Parameters}
    \label{parameter}
    \begin{threeparttable}
    \begin{tabular}{c|c}
        \hline
        Parameter &  Value\\
        \hline
        Number of orbital planes & 6\\
        Number of satellites per orbital plane & 8\\
        Orbital plane inclination & $40^{\circ}$\\
        Orbital height & 1200 km (LEO) \cite{3GPP-38821}\\
        Number of sub-arrays & $N_{x}^\text{sub}$= 8,$N_{y}^\text{sub}$= 4\\
        Number of antenna elements per sub-array & $N_x$ = $N_y$ = 8\\
        Number of GUs & $N_u = 80$\\
        \hline
        Downlink carrier frequency & $20$ GHz \cite{3GPP-38821}\\
        Bandwidth & 400 MHz \cite{3GPP-38821}\\
        Receiver noise temperature & 24 dBK \cite{hybrid}\\
        Satellite antenna gain\tnote{*} & 21.5 dBi\\
        VSAT maximum antenna gain & 40 dBi \cite{3GPP-38821}\\
        Transmission power per satellite & $P_\text{T}=80$ W\\
        \hline
    \end{tabular}
    \begin{tablenotes}
        \item[*] The satellite antenna is a kind of phased array antenna, whose gain can be calculated according to the number and size of antenna elements and the carrier wavelength.
    \end{tablenotes}
    \end{threeparttable}
\end{table}

We consider a $6 \times 8$ LEO Walker constellation with an inclination of $40^{\circ}$ as an example, which has the ability to achieve full-time coverage of the entire China according to the simulation result of a popular aerospace simulation software Systems Tool Kit (STK) \cite{STK}. 
We use STK to simulate the movement of the constellation and we sample every 20 minutes within eight hours from 4am to 12pm in Beijing time, resulting in 24 experiments to test the performances of the proposed algorithms and schemes. 
Besides, all GUs are located in suburban areas of 80 representative cities in China. The simulation parameters are given in Table \ref{parameter}. Note that the proposed algorithm and main observations still hold for other types of LEO constellations.

The baseline scheme is analog beamforming and user scheduling scheme, denoted as AU. The specific steps of AU scheme are: (1) analog beamforming as described in Section \ref{sec-3A}, (2) user scheduling using Algorithm \ref{PUS}, and (3) power scaling.
\begin{figure*}[htb]
    \centering
    \subfigure[Beijing]{
        \centering
        \includegraphics[width=0.17\textwidth]{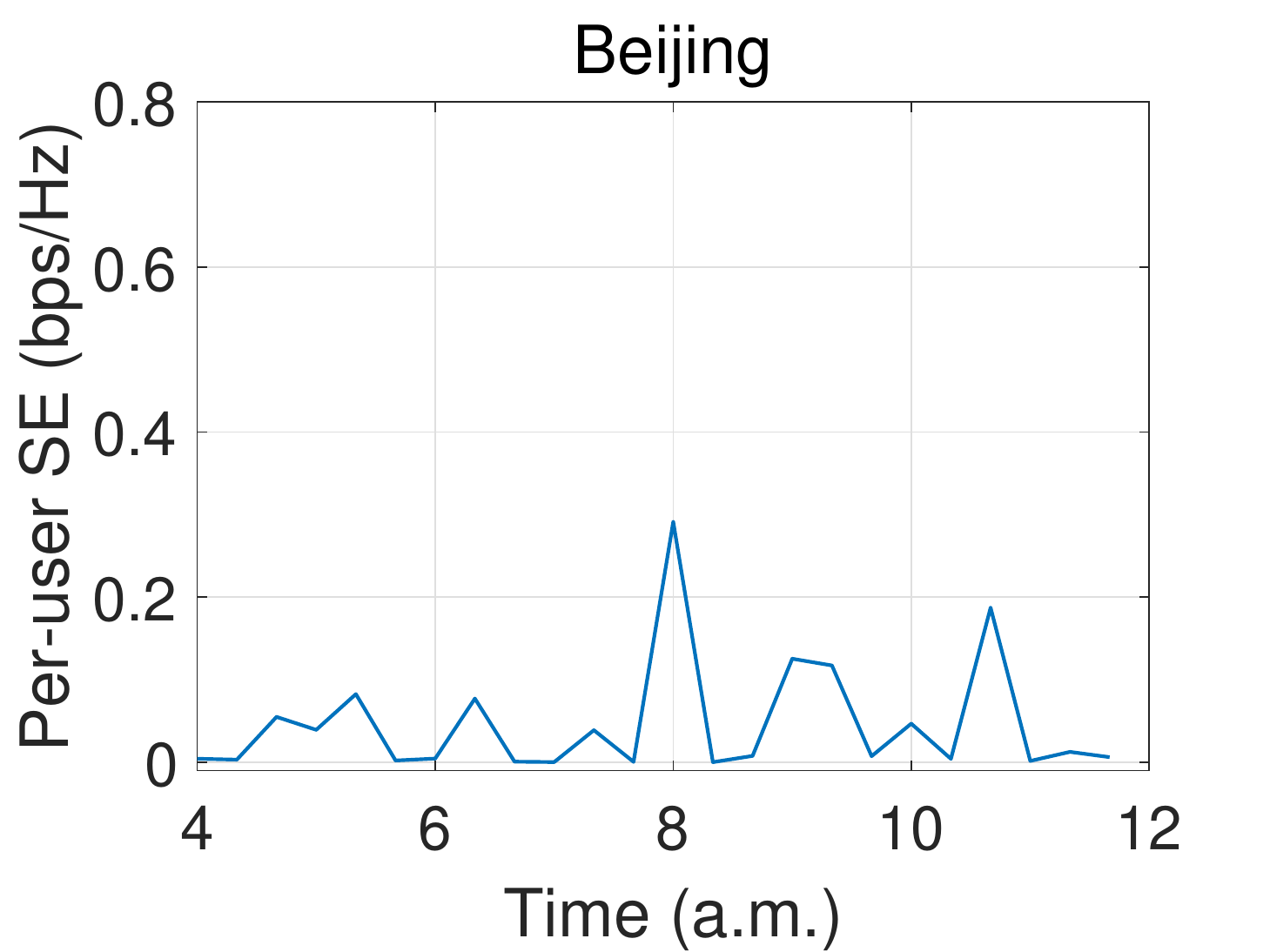}
    }
    \subfigure[Shanghai]{
        \centering
        \includegraphics[width=0.17\textwidth]{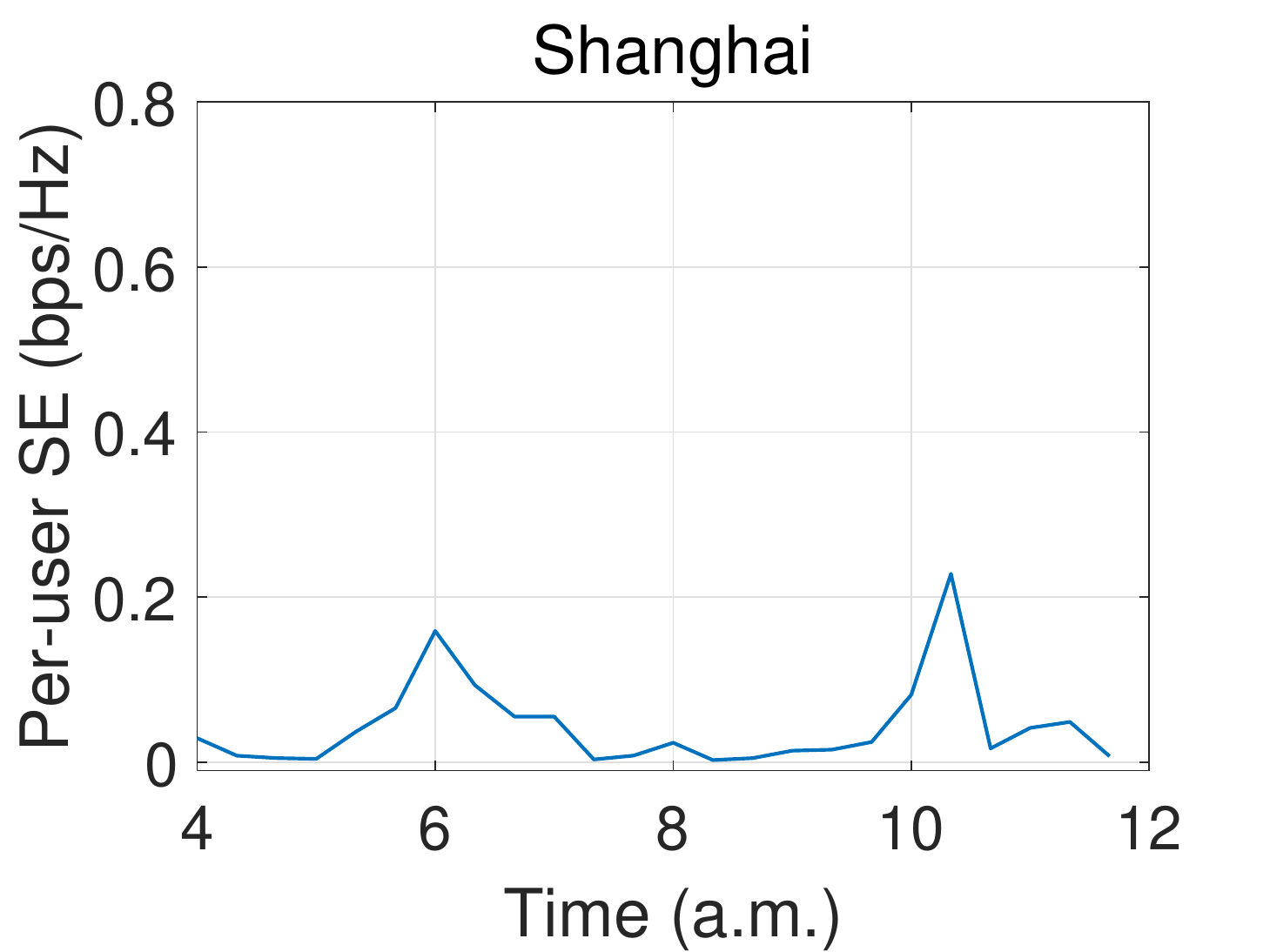}
    }
    \subfigure[Wuhan]{
        \centering
        \includegraphics[width=0.17\textwidth]{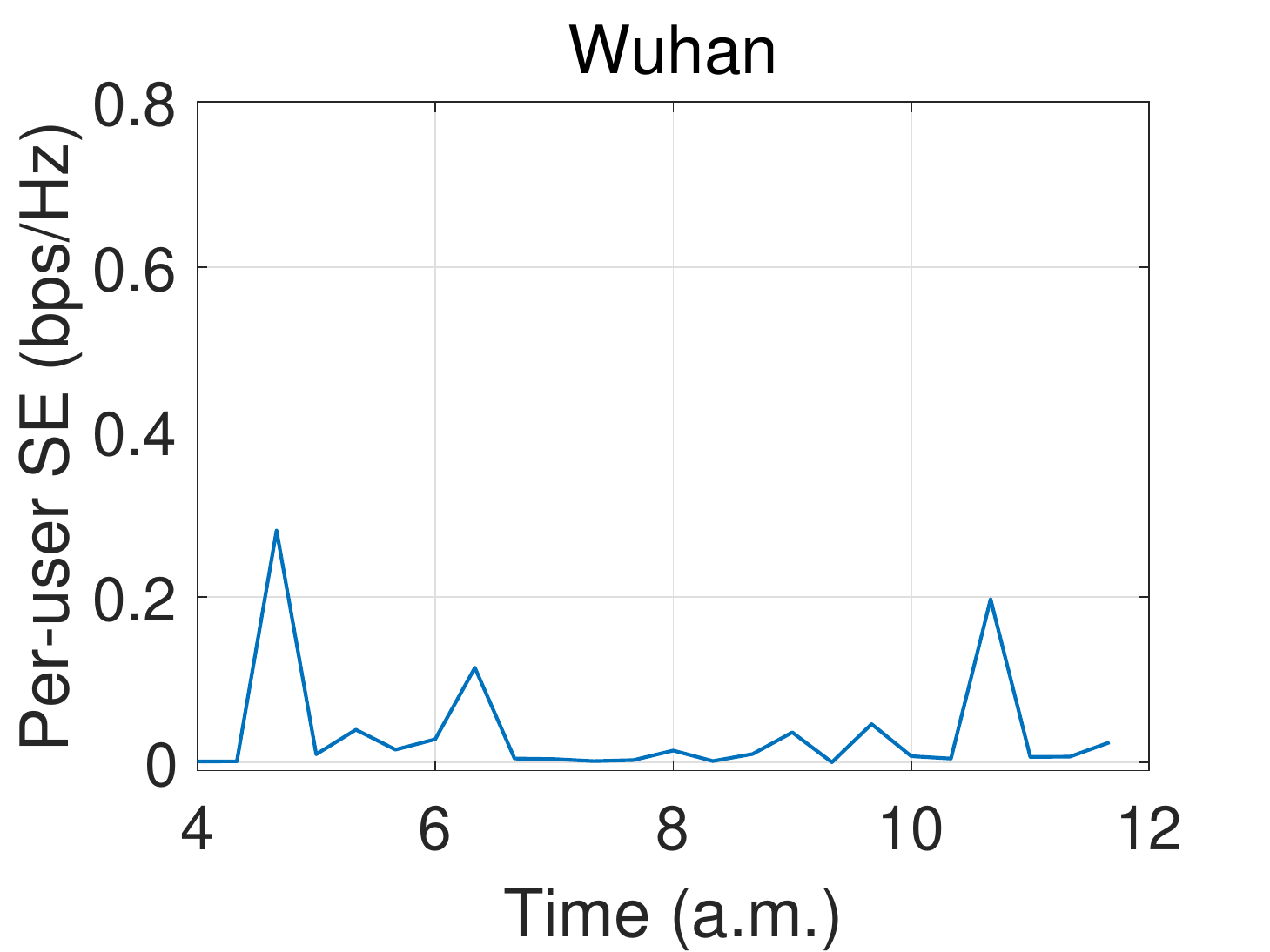}
    }
    \subfigure[Kashi]{
        \centering
        \includegraphics[width=0.17\textwidth]{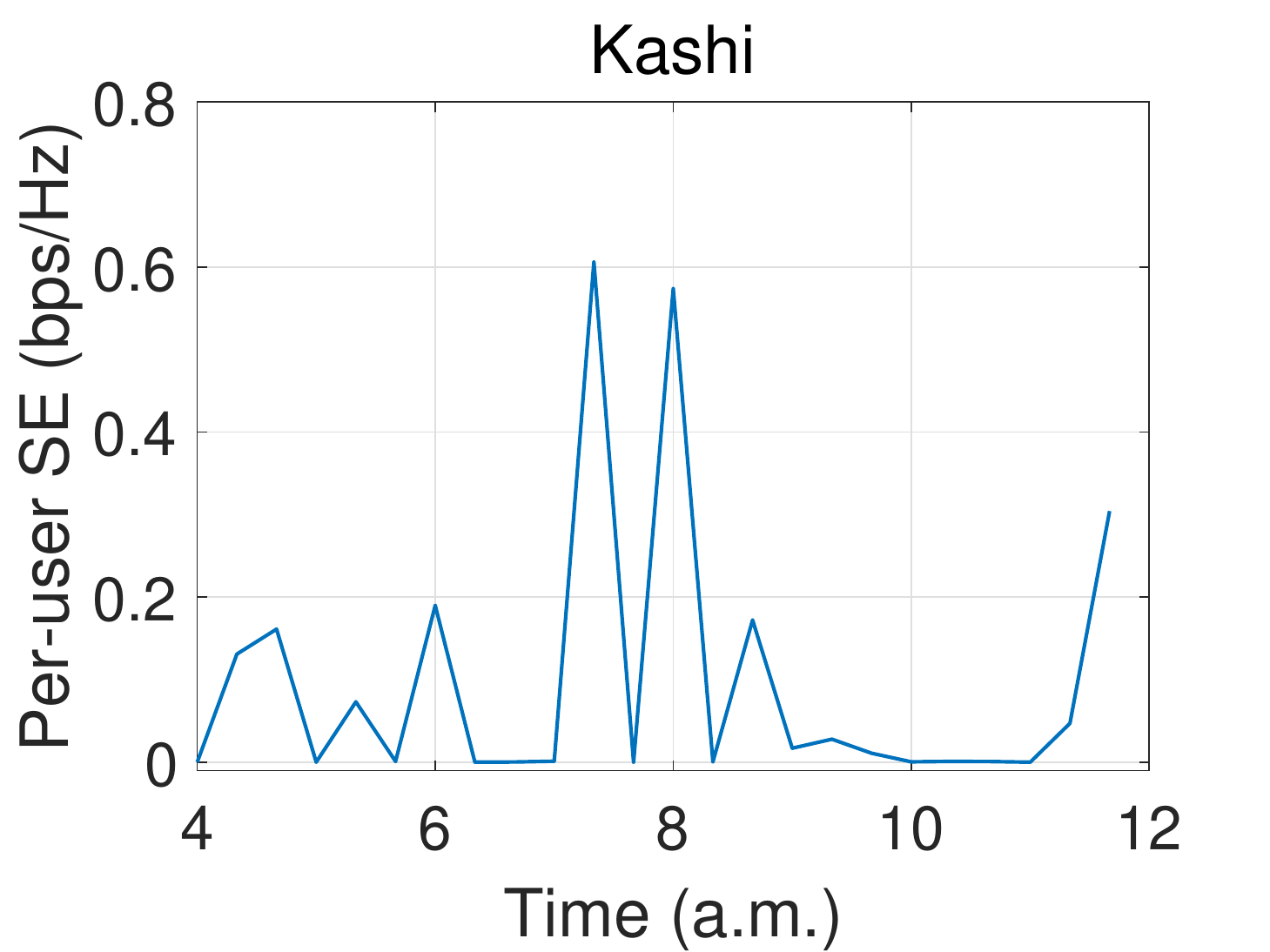}
    }
    \subfigure[Nansha]{
        \centering
        \includegraphics[width=0.17\textwidth]{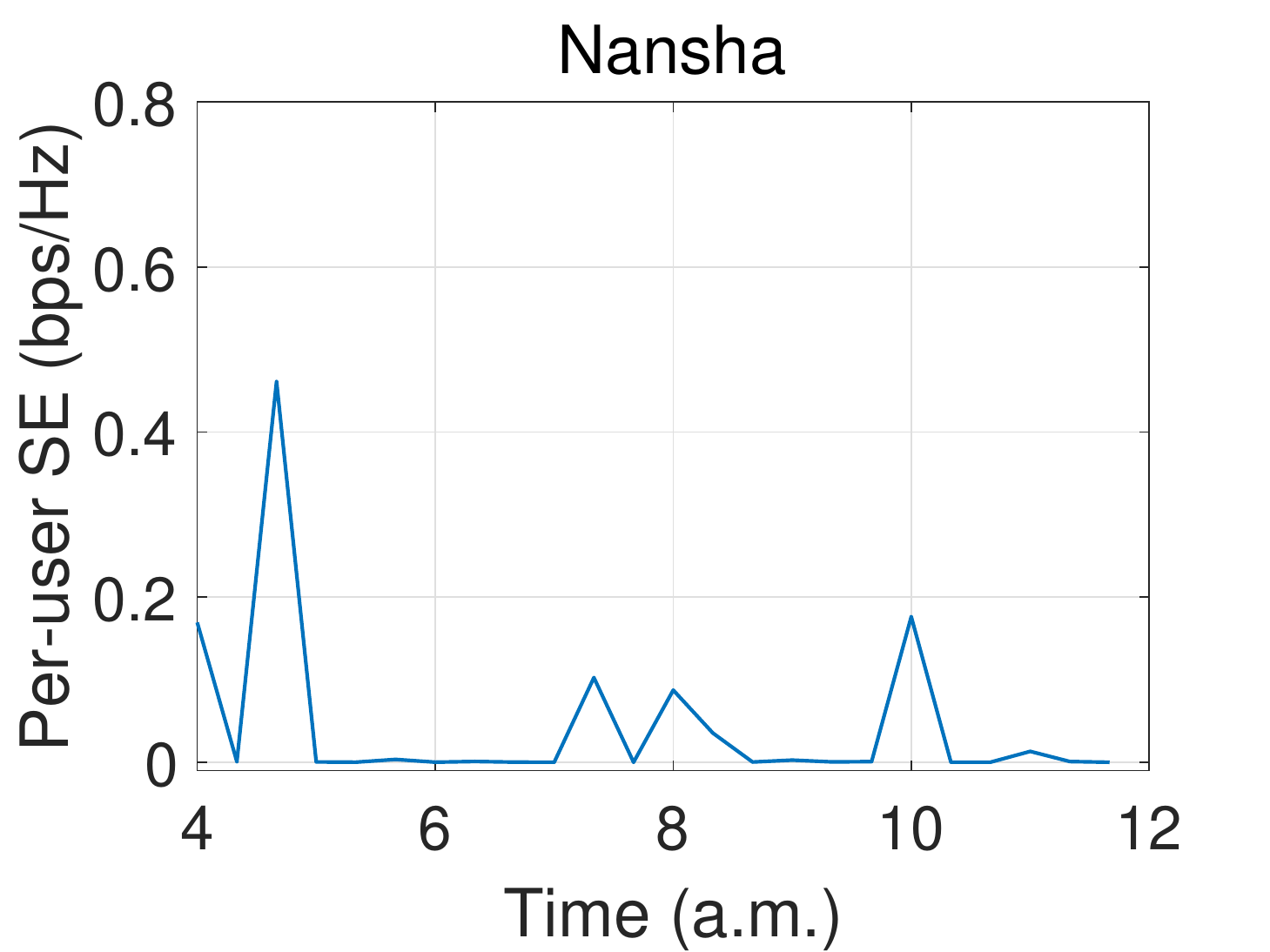}
    }
    \caption{Per-user SEs of GUs in five typical cities.}
    \label{5cities}
    \vspace{-0.4cm}
\end{figure*}
 The SE performances of AU, SHU and JHU are shown in Fig. \ref{mean-sum}, where the left picture illustrates the total SEs in 24 experiments and the right picture shows the corresponding mean total SEs averaged over the 24 experiments. It can be seen that AU performs the worst because of the lack of interference mitigation. By performing digital beamforming after user scheduling to mitigate interference, the performance of SHU increases by 10.2\% compared with AU on average. However, SHU cannot make full use of link information since digital beamforming is implemented independently of user scheduling. In JHU, the digital beamforming matrix is updated in real time when calculating the total SE increment and establishing links, leading to increase of SE by 47.2\% compared with SHU and 62.3\% compared with AU on average.

From Fig. \ref{mean-sum}, we can observe that the network total SE fluctuates significantly with time. The reason is that the topological relationships between satellites and GUs change rapidly with time. Taking GUs in five typical cities as examples, we calculate their SEs with JHU and show the results in Fig. \ref{5cities}. These five cities lie in the northern, eastern, middle, western, and southern part of China, respectively. The SEs of GUs in these five cities fluctuate differently due to their different geographical positions and topological relationships. For the GU in each city, the SE varies with time because of the change of topological relationship, which is caused by satellites' movement. These five GUs have very different SEs at the same time because of different sets of visible satellites for these GUs and different elevation angles and distances between satellites and GUs, leading to different path losses. The different sets of visible satellites and path losses are both caused by discrepant geographical positions of these GUs. 

\begin{table}[tb]
    \centering
    \caption{SE statistics of dense and sparse GUs}
    \begin{tabular}{c|c|c}
        \hline
          & Mean Value of SEs (bps/Hz) & Variance of SEs (bps/Hz)$^2$\\
        \hline
        Dense GUs & 0.028569 & 0.035576\\
        Sparse GUs & 0.049798 & 0.064577\\
        \hline
    \end{tabular}
    \label{mean-var}
    \vspace{-0.5cm}
\end{table}

In Fig. \ref{5cities}, we can also find that the maximum SEs of GUs in Kashi and Nansha are higher than those in other cities. Part of the reason lies in the density of GUs. Sparse GUs refer to the users in an area where inter-GU distances are all greater than $D$. The opposite holds for dense GUs. In our study, we set $D$ to 400km as an example. Among these 80 GUs, there are 12 sparse GUs and 68 dense GUs. Correspondingly, the GUs in Kashi and Nansha are sparse GUs and the other three GUs are dense GUs. In order to obtain more comprehensive observations, we calculate the SEs of all dense and sparse GUs in 24 experiments and obtain their respective mean value and variance in Table \ref{mean-var}.
The mean SE of sparse GUs are nearly twice as much as that of dense GUs. The potential reasons are: The sparse GUs are surrounded by fewer GUs and suffer less interference from others hence having higher SINR; Most of the sparse GUs are located in the border areas of China, thus they have greater probability of monopolizing one satellite and getting larger transmission power. Additionally, the variance of SE of sparse GUs is much larger than that of dense GUs, which indicates that the SEs of sparse GUs fluctuate more substantially than dense GUs due to their geographical positions.

\section{Conclusion}
In this paper, we first provide a hybrid beamforming architecture of UPAs which is suitable for satellites because of the limitation of on-board RF chains. 
According to the hybrid architecture, we introduce an analog beamforming method based on the 2D DFT codebook which generates a desired codeword by linearly combining four selected codewords. 
Then digital beamforming in accordance with regularized ZF is employed to mitigate the inter-beam interference and scale the transmission power. 
Moreover, we propose a heuristic user scheduling algorithm to determine the links between satellites and GUs. Subsequently, we propound two implementation schemes: a separate scheme and a joint scheme. 
Simulation results show that the JHU scheme outperforms SHU because of the joint design of beamforming and user scheduling, and can achieve an average gain of 47.2\% compared with SHU. 
Furthermore, based on the simulation results, we analyze the key factors influencing GUs' SEs, including geographical positions, topological relationships, and the density of GUs. 
Future work may consider the impact of the satellite constellation design on the network's communication performance and different beamforming architectures or algorithms.


\end{document}